\documentclass[a4paper,fleqn,usenatbib]{mnras}

\usepackage{newtxtext,newtxmath}

\usepackage[T1]{fontenc}
\usepackage{ae,aecompl}

\usepackage{graphicx}	
\usepackage{amsmath}	
\usepackage{rotating}

\newcommand{\Eexc}{$E_{\rm exc}$}
\newcommand{\Teff}{T_{\rm eff}}
\newcommand{\logg}{\rm log~ g}

\newcommand{\eps}[1]{\log\varepsilon_{\rm #1}}
\newcommand{\kH}{$S_{\rm H}$}    
\newcommand{\eu}[5]{\mbox{$#1\,^#2{\rm #3}^{#4}_{\rm #5}$}}
\def\ione{\,{\sc i}}
\def\ii{\,{\sc ii}}

\title[Non-LTE abundance corrections for metal-poor stars]{1D non-LTE corrections for chemical abundance analyses \\ of very metal-poor stars }

\author[L. Mashonkina et al.]{L.~Mashonkina,$^{1,2}$\thanks{E-mail: lima@inasan.ru}
Yu.~Pakhomov,$^{1}$
T.~Sitnova,$^{1}$ 
A.~Smogorzhevskii,$^{1,3}$
\newauthor P.~Jablonka,$^{4,5}$
and V.~Hill$^{6}$ \\
$^{1}$Institute of Astronomy of the Russian Academy of Sciences, Pyatnitskaya st. 48, 119017, Moscow, Russia \\
$^{2}$ Institute of Laser Physics, SB RAS, Ac. Lavrentieva ave. 13, 630090, Novosibirsk, Russia \\
$^{3}$ M. V. Lomonosov Moscow State University, Kolmogorova st. 1, 119991, Moscow, Russia \\
$^{4}$ Laboratoire d' Astrophysique, Ecole Polytechnique F\'ed\'erale de Lausanne (EPFL), Observatoire de Sauverny, CH-1290 Versoix, Switzerland \\
$^{5}$ GEPI, Observatoire de Paris, CNRS, Universit\'e Paris Diderot, F-92125 Meudon Cedex, France \\
$^{6}$ Laboratoire Lagrange, Universit\'e de Nice Sophia-Antipolis, Observatoire de la C\^ote d'Azur, France
}
\date{Accepted XXX. Received YYY; in original form ZZZ}

\pubyear{2021}

\begin{document}
\label{firstpage}
\pagerange{\pageref{firstpage}--\pageref{lastpage}}
\maketitle

\begin{abstract}
Detailed chemical abundances of very metal-poor (VMP, [Fe/H] $< -2$) stars are important for better understanding the First Stars, early star formation and chemical enrichment of galaxies. Big on-going and coming high-resolution spectroscopic surveys provide a wealth of material that needs to be carefully analysed. For VMP stars, their elemental abundances should be derived based on the non-local thermodynamic equilibrium (non-LTE = NLTE) line formation because low metal abundances and low electron number density in the atmosphere produce the physical conditions favorable for the departures from LTE. The galactic archaeology research requires homogeneous determinations of chemical abundances. For this purpose, we present grids of the 1D-NLTE abundance corrections for lines of Na\ione, Mg\ione, Ca\ione, Ca\ii, Ti\ii, Fe\ione, Zn\ione, Zn\ii, Sr\ii, and Ba\ii\ in the range of atmospheric parameters that represent VMP stars on various evolutionary stages and cover effective temperatures from 4000 to 6500~K, surface gravities from $\logg$ = 0.5 to $\logg$ = 5.0, and metallicities $-5.0 \le$ [Fe/H] $\le -2.0$. The data is publicly available, and we provide the tools for interpolating in the grids online.
\end{abstract}

\begin{keywords}
line: formation -- stars: abundances -- stars: atmospheres.
\end{keywords}


\section{Introduction}

Very metal-poor (VMP, [Fe/H]\footnote{In the classical notation, where [X/H] = $\log(N_{\rm X}/N_{\rm
    H})_{star} - \log(N_{\rm X}/N_{\rm H})_{\odot}$.} $< -2$) stars are fossils of the early epochs of star formation in their parent galaxy.
Their detailed elemental abundances are of extreme importance for understanding the nature of the First Stars, uncovering the initial mass function and the metallicity distribution function of the galaxy, testing the nuclesynthesis theory predictions and the galactic chemical evolution models \citep{2005ARA&A..43..531B,2015ARA&A..53..631F,2019PhT....72b..36W,2020ApJ...900..179K}.
Since 1980th, the number of discovered VMP star candidates has grown tremendously thanks to the wide-angle spectroscopic and photometric surveys, such as HK \citep{1985AJ.....90.2089B,1992AJ....103.1987B}, HES \citep[Hamburg/ESO,][]{2002A&A...391..397C}, RAVE \citep{2006AJ....132.1645S}, SMSS \citep[SkyMapper Southern Sky,][]{2007PASA...24....1K}, SEGUE/SDSS \citep{2009AJ....137.4377Y}, LAMOST \citep{2012RAA....12..735D}.
    The survey Pristine has been specially designed for efficient searching VMP stars \citep{2017MNRAS.471.2587S}. Using the narrow-band photometric filter centered on the Ca\ii\ H \& K lines makes possible to successfully predict stellar metallicities \citep{2017MNRAS.472.2963Y,2020MNRAS.492.3241V}. 

The number of confirmed VMP stars is substantially lower than the number of candidates because the verification of very low metallicity  requires the high-resolution follow-ups. The SAGA (Stellar Abundances for Galactic Archaeology) database \citep{2008PASJ...60.1159S} includes about 1390 Galactic stars with [Fe/H] $\le -2$, for which their metallicities were derived from the R = $\lambda /\Delta \lambda \ge 20\,000$ spectra. The 470 stars of them have [Fe/H] $\le -3$, and 28 stars are ultra metal-poor (UMP, [Fe/H] $\le -4$). A burst in the number of VMP stars with detailed elemental abundances derived is expected with the launch of the WEAVE (WHT Enhanced Area Velocity Explorer) project \citep[see a description in][science observations start in November of 2022]{2014SPIE.9147E..0LD}. A vast amount of spectral data will be taken with the coming 4-metre Multi-Object Spectroscopic Telescope \citep[4MOST,][]{2019Msngr.175....3D}.

 Abundance ratios among the elements of different origin, such as Mg and Fe, for stellar samples covering broad metallicity ranges serve as the observational material for the galactic archaeology research.
The simplest and widely applied method to derive elemental abundances is based on using one-dimensional (1D) model atmospheres and the assumption of local thermodynamic equilibrium (LTE), see, for example, the abundance results from the high-resolution spectroscopic survey APOGEE \citep{2017AJ....154...94M,2020ApJS..249....3A}.
In metal-poor atmospheres, in particular, of cool giants, low total gas pressure and low electron number density lead to departures from LTE that grow towards lower metallicity due to decreasing collisional rates and increasing radiative rates as a result of dropping ultra-violet (UV) opacity. The non-local thermodynamic equilibrium (non-LTE = NLTE) line formation calculations show that the NLTE effects for lines of one chemical species and for different chemical species are different in magnitude and sign, depending on the stellar parameters and element abundances. Ignoring the NLTE effects leads to a distorted picture of the galactic abundance trends and thus to wrong conclusions about the galactic chemical evolution.

The NLTE abundance from a given line in a given star can be obtained by adding the theoretical NLTE abundance correction, which corresponds to the star's atmospheric parameters, to the LTE abundance derived from the observed spectrum: $\eps{NLTE} = \eps{LTE} + \Delta_{\rm NLTE}$. For a number of chemical species, $\Delta_{\rm NLTE}$ can be taken online from the websites
\begin{itemize}
\item INSPECT ({\url{http://www.inspect-stars.com}}) for lines of Li\ione, Na\ione, Mg\ione, Ti\ione, Fe\ione -\ii, and Sr\ii,
\item NLTE\_MPIA ({\url{http://nlte.mpia.de/}}) for lines of O\ione, Mg\ione, Si\ione, Ca\ione -\ii, Ti\ione -\ii, Cr\ione, Mn\ione, Fe\ione -\ii, and Co\ione,
\item {\url{http://spectrum.inasan.ru/nLTE/}} for lines of Ca\ione, Ti\ione -\ii, and Fe\ione.
\end{itemize}

Extensive grids of the NLTE abundance corrections are provided by \citet[][lines of Ba\ii]{2015A&A...581A..70K}, \citet[][K\ione]{2019A&A...627A.177R}, and \citet[][Na\ione, Mg\ione, Al\ione]{2022A&A...665A..33L}.
The NLTE abundance corrections for the selected lines of S\ione\ and Zn\ione\ in the limited set of atmospheric models were computed by \citet{Takeda2005zn}. \citet{2011MNRAS.418..863M} report the NLTE to LTE equivalent width ratios for lines of Mg\ione, Ca\ione, and Ca\ii\ in the grid of model atmospheres representing cool giants.

Different approach is a determination of the NLTE abundance directly, by using the synthetic spectrum method and the precomputed departure coefficients, b$_i$ = $n_i^{\rm NLTE}/n_i^{\rm LTE}$, for the chemical species under investigation. Here, $n_i^{\rm NLTE}$ and $n_i^{\rm LTE}$ are the statistical equilibrium and the Saha-Boltzmann's number densities, respectively, for the enery level $i$.
\citet{2020A&A...642A..62A} provide the grids of b$_i$
for 13 chemical species (neutral H, Li, C, N, O, Na, Mg, Al, Si, K, Ca, Mn; singly ionized Mn, Ba)
across a grid of the classical one-dimensional (1D) MARCS model atmospheres \citep{Gustafssonetal:2008}.
This approach is based on using the 1D-NLTE spectral synthesis codes, such as {\sc SME} \citep{2017A&A...597A..16P} {\sc synthV}\_NLTE \citep{2019ASPC}, Turbospectrum \citep{2023A&A...669A..43G}.

An approach based on three-dimensional (3D) model atmospheres combined with the NLTE line formation is extremely time consuming and, to now, was applied to a few chemical species in the Sun \citep{2015A&A...583A..57S,2017MNRAS.468.4311L,2017MNRAS.464..264A,2018A&A...616A..89A,2019A&A...624A.111A,2020A&A...636A.120A,2020A&A...634A..55G} and the benchmark VMP stars \citep{2016MNRAS.463.1518A,2017A&A...597A...6N,2019A&A...631A..80B}. Grids of the 3D-NLTE abundance corrections were computed for lines of O\ione\ \citep{2016MNRAS.455.3735A} and Fe\ione -\ii\ \citep{2022A&A...668A..68A} using the STAGGER grid of model atmospheres for a limited range of effective temperatures ($\Teff$ = 5000-6500~K), surface gravities ($\logg$ = 3.0-4.5), and metallicities ([Fe/H] = 0 to $-3$). For the Li\ione\ lines, grids of the 3D-NLTE abundance corrections were computed by \citet{2020A&A...638A..58M} and \citet{2021MNRAS.500.2159W} with the CO$^5$BOLD and STAGGER model atmospheres, respectively.

The 3D-NLTE calculations are available for a small number of the chemical elements observed in VMP stars, and they cover only in part the range of relevant atmospheric parameters. Furthermore, as shown by \citet{2022A&A...668A..68A} for Fe\ione, the abundance differences between 3D-NLTE and 1D-NLTE are generally less severe compared with the differences between 3D-NLTE and 1D-LTE and reach 0.2~dex, at maximum (see Figs.~5-7 in their paper). Therefore, calculations of the 1D-NLTE abundance corrections for extended linelists across the stellar parameter range which represents the VMP stars make sense, and they are useful for the galactic archaeology research. Availability and comparison of $\Delta_{\rm NLTE}$ from different independent studies increase a credit of confidence in the spectroscopic NLTE analyses.

This paper presents the 1D-NLTE abundance corrections for lines of 10 chemical species in the grid of MARCS model atmospheres with $\Teff$ = 4000-6500~K, $\logg$ = 0.5-5.0, and $-5 \le$ [Fe/H] $\le -2$.
We provide the tools for calculating online the NLTE abundance correction(s) for  given line(s) and given atmospheric parameters by interpolating in the precomputed grids.

Potential users may take the following advantages of our data compared with the grids of the 1D-NLTE abundance corrections available in the literature.
\begin{itemize}
\item Only this study provides extended grids of the NLTE abundance corrections for lines of  Zn\ii\ and Ba\ii.
\item For Ca\ione\ and Ca\ii, the NLTE calculations were performed with advanced treatment of the Ca\ione\ + H\ione\ and Ca\ii\ + H\ione\ collisions, following \citet{2017ApJ...851...59B} and \citet{2019PhRvA.100f2710B}, respectively.
\item For Zn\ione\ and Sr\ii, our results are based on advanced treatment of collisions with  H\ione, following \citet[][Zn\ione]{2022MNRAS.515.1510S} and \citet[][Sr\ii]{2022arXiv220308629Y}. Our grids cover the broader range of $\Teff$, $\logg$, and [Fe/H] compared to that for Zn\ione\ in \citet{Takeda2005zn} and for Sr\ii\ in the INSPECT database.
\item For Ca\ione --Ca\ii, Fe\ione --Fe\ii, and Na\ione, the developed 1D-NLTE methods have been verified with spectroscopic analyses of VMP stars and have been shown to yield reliable results.
\end{itemize}

The paper is organised as follows. Section~\ref{Sect:NLTE} describes our NLTE methods and their verification with observations of VMP stars. New grids of the NLTE abundance corrections are presented in Sect.~\ref{Sect:grids}. In Sect.~\ref{Sect:others}, we compare our calculations with those from other studies. Our recommendations and final remarks are given in Sect.~\ref{Sect:conclusion}.


\section{NLTE methods and their verification}\label{Sect:NLTE}

The present investigation is based on the NLTE methods developed and tested in our earlier studies.
Details of the adopted atomic data and the NLTE line formation for Na\ione, Mg\ione, Ca\ione -\ii, Ti\ione -Ti\ii, Fe\ione -\ii, Zn\ione -\ii, Sr\ii, and Ba\ii\ can be found in the papers cited in Table~\ref{Tab:nlte}.
It is important to note that collisions with hydrogen atoms were treated with the data based on quantum-mechanical calculations.
The exceptions are Ti\ii\ and Fe\ione -\ii, for which we adopted the Drawinian rates \citep{Drawin1969,Steenbock1984} scaled by an empirically estimated factor of \kH\ = 1 \citep{sitnova_ti} and \kH\ = 0.5 \citep{2015ApJ...808..148S,dsph_parameters}, respectively.

\begin{table} 
 \caption{\label{Tab:nlte} Atomic models used in this study.}
 \begin{tabular}{lll}\hline\hline \noalign{\smallskip}
 Species & Reference & H\ione\ collisions  \\
\noalign{\smallskip} \hline \noalign{\smallskip}
Na\ione    &   \citet{alexeeva_na} &  BBD10 \\
Mg\ione    &   \citet{mash_mg13}   &  BBS12 \\
Ca\ione -\ii & \citet{2017AA...605A..53M}, & BVY17 \\
             & \citet{2020AstL...46..621N} & BVY19 \\
Ti\ii   &  \citet{sitnova_ti}    &  \kH\,= 1 \\
Fe\ione -\ii & \citet{mash_fe}       &  \kH\,= 0.5 \\
Zn\ione -\ii & \citet{2022MNRAS.515.1510S} & YB22 \\
Sr\ii   &   \citet{2022MNRAS.509.3626M} & YBM22 \\
Ba\ii   &   \citet{2019AstL...45..341M} &  BY18 \\
\noalign{\smallskip}\hline \noalign{\smallskip}
\end{tabular}

{\bf Notes.} Collisions with H\ione\ are treated following to
BBD10 = \citet{barklem2010_na},
BBS12 = \citet{mg_hyd2012},
BVY17 = \citet[][Ca\ione]{2017ApJ...851...59B},
BVY19 = \citet[][Ca\ii]{2019PhRvA.100f2710B},
BY18 = \citet{2018MNRAS.478.3952B},
YB22 = Yakovleva S.~A. and Belyaev A.~K., as presented in \citet{2022MNRAS.515.1510S},
YBM22 = \citet{2022arXiv220308629Y},
and \citet{Steenbock1984}, with using the scaling factor \kH.
\end{table}

The code {\sc detail} \citep{Giddings81,Butler84} with the revised opacity package \citep[see the description in][]{mash_fe} was used to solve the coupled radiative transfer and statistical equilibrium (SE) equations. The obtained LTE and NLTE level populations were then implemented in the code {\sc linec} \citep{Sakhibullin1983} that, for each given spectral line, computes the NLTE curve of growth and finds the shift in the NLTE abundance, which is required to reproduce the LTE equivalent width. Such an abundance shift is referred to as the NLTE abundance correction, $\Delta_{\rm NLTE} = \eps{NLTE}-\eps{LTE}$.

All the calculations were performed using the classical LTE model atmospheres with the standard chemical composition \citep{Gustafssonetal:2008}, as provided by the MARCS website\footnote{\url{http://marcs.astro.uu.se}}.


Below we provide evidence for a correct treatment of the NLTE line formation for Fe\ione -Fe\ii, Ca\ione -Ca\ii, and Na\ione\ in the atmospheres of VMP stars.

\begin{table} 
 \centering
 \caption{\label{Tab:gaia} Distance-based and spectroscopic surface gravities for the Galactic stellar samples from \citet{2015ApJ...808..148S} and \citet{dsph_parameters}.}
 \begin{tabular}{lccccr}
\hline\hline \noalign{\smallskip}
 \multicolumn{1}{c}{Star} & \multicolumn{2}{c}{Gaia eDR3} & $\Teff$ & $\logg$ & [Fe/H] \\
\cline{2-3} 
             &   d(pc)  & $\logg_{\rm d}$ &  (K) &      (Sp)     &        \\
\noalign{\smallskip} \hline 
\multicolumn{1}{c}{1} & \multicolumn{1}{c}{2} & 3 &  4      &  5   & \multicolumn{1}{c}{6} \\      
 \hline \noalign{\smallskip}
HD~2796      &  631.3 \scriptsize{(9.3) }&  1.79 (0.03) & 4880 &  1.55 & --2.32 \\
HD~4306      &  480.6 \scriptsize{(7.7) }&  2.26 (0.03) & 4960 &  2.18 & --2.74 \\
HD~8724      &  425.8 \scriptsize{(4.6) }&  1.77 (0.03) & 4560 &  1.29 & --1.76 \\
HD~19373     &  10.57 \scriptsize{(0.02)} & 4.18 (0.02) & 6045 &  4.24 &  0.10 \\
HD~22484     &  13.91 \scriptsize{(0.03)} & 4.03 (0.02) & 6000 &  4.07 &  0.01 \\
HD~22879     &  26.07 \scriptsize{(0.02)} & 4.23 (0.02) & 5800 &  4.29 & --0.84 \\
HD~24289     &  211.9 \scriptsize{(1.9) }&  3.79 (0.03) & 5980 &  3.71 & --1.94 \\
HD~30562     &  26.12 \scriptsize{(0.03)} & 4.09 (0.02) & 5900 &  4.08 &  0.17 \\
HD~30743     &  36.22 \scriptsize{(0.03)} & 4.13 (0.02) & 6450 &  4.20 & --0.44 \\
HD~34411     &  12.55 \scriptsize{(0.02)} & 4.23 (0.02) & 5850 &  4.23 &  0.01 \\
HD~43318     &  36.30 \scriptsize{(0.09)} & 3.90 (0.02) & 6250 &  3.92 & --0.19 \\
HD~45067     &  32.85 \scriptsize{(0.05)} & 3.96 (0.02) & 5960 &  3.94 & --0.16 \\
HD~45205     &  75.59 \scriptsize{(0.12)} & 3.85 (0.02) & 5790 &  4.08 & --0.87 \\
HD~49933     &  29.79 \scriptsize{(0.04)} & 4.17 (0.02) & 6600 &  4.15 & --0.47 \\
HD~52711     &  18.93 \scriptsize{(0.02)} & 4.32 (0.02) & 5900 &  4.33 & --0.21 \\
HD~58855     &  20.38 \scriptsize{(0.06)} & 4.28 (0.02) & 6410 &  4.32 & --0.29 \\
HD~59374     &  57.86 \scriptsize{(0.06)} & 4.29 (0.02) & 5850 &  4.38 & --0.88 \\
HD~59984     &  28.64 \scriptsize{(0.03)} & 3.96 (0.02) & 5930 &  4.02 & --0.69 \\
HD~62301     &  34.07 \scriptsize{(0.03)} & 4.09 (0.02) & 5840 &  4.09 & --0.70 \\
HD~64090     &  27.32 \scriptsize{(0.02)} & 4.60 (0.03) & 5400 &  4.70 & --1.73 \\
HD~69897     &  18.21 \scriptsize{(0.03)} & 4.24 (0.02) & 6240 &  4.24 & --0.25 \\
HD~74000     &  109.8 \scriptsize{(0.2) } &  4.27 (0.02) & 6225 &  4.13 & --1.97 \\
HD~76932     &  21.41 \scriptsize{(0.02)} & 4.09 (0.02) & 5870 &  4.10 & --0.98 \\
HD~82943     &  27.66 \scriptsize{(0.02)} & 4.37 (0.02) & 5970 &  4.37 &  0.19 \\
HD~84937     &  73.87 \scriptsize{(0.24)} & 4.15 (0.02) & 6350 &  4.09 & --2.16 \\
HD~89744     &  38.50 \scriptsize{(0.06)} & 3.98 (0.02) & 6280 &  3.97 &  0.13 \\
HD~90839     &  12.94 \scriptsize{(0.01)} & 4.33 (0.02) & 6195 &  4.38 & --0.18 \\
HD~92855     &  36.63 \scriptsize{(0.03)} & 4.40 (0.02) & 6020 &  4.36 & --0.12 \\
HD~94028     &  49.39 \scriptsize{(0.06)} & 4.33 (0.02) & 5970 &  4.33 & --1.47 \\
HD~99984     &  51.50 \scriptsize{(0.10)} & 3.72 (0.02) & 6190 &  3.72 & --0.38 \\
HD~100563    &  26.86 \scriptsize{(0.04)} & 4.25 (0.02) & 6460 &  4.32 &  0.06 \\
HD~102870    &  11.00 \scriptsize{(0.02)} & 4.06 (0.02) & 6170 &  4.14 &  0.11 \\
HD~103095    &   9.17 \scriptsize{(0.00)} & 4.66 (0.03) & 5130 &  4.66 & --1.26 \\
HD~105755    &  83.31 \scriptsize{(0.10)} & 4.05 (0.02) & 5800 &  4.05 & --0.73 \\
HD~106516    &  22.32 \scriptsize{(0.40)} & 4.38 (0.03) & 6300 &  4.44 & --0.73 \\
HD~108177    &  105.9 \scriptsize{(0.2) } &  4.25 (0.02) & 6100 &  4.22 & --1.67 \\
HD~108317    &  192.7 \scriptsize{(1.1) } &  2.81 (0.03) & 5270 &  2.96 & --2.18 \\
HD~110897    &  17.55 \scriptsize{(0.01)} & 4.36 (0.02) & 5920 &  4.41 & --0.57 \\
HD~114710    &   9.19 \scriptsize{(0.01)} & 4.45 (0.02) & 6090 &  4.47 &  0.06 \\
HD~115617    &   8.53 \scriptsize{(0.01)} & 4.35 (0.03) & 5490 &  4.40 & --0.10 \\
HD~122563    &  318.6 \scriptsize{(3.4) } & 1.32 (0.03) & 4600 &  1.32 & --2.63 \\
HD~128279    &  130.4 \scriptsize{(0.5) } & 3.01 (0.03) & 5200 &  3.00 & --2.19 \\
HD~134088    &  39.23 \scriptsize{(0.04)} & 4.36 (0.02) & 5730 &  4.46 & --0.80 \\
HD~134169    &  53.88 \scriptsize{(0.39)} & 4.09 (0.02) & 5890 &  4.02 & --0.78 \\
HD~138776    &  76.11 \scriptsize{(0.15)} & 4.17 (0.03) & 5650 &  4.30 &  0.24 \\
HD~140283    &  61.32 \scriptsize{(0.10)} & 3.73 (0.02) & 5780 &  3.70 & --2.46 \\
HD~142091    &  29.96 \scriptsize{(0.07)} & 3.10 (0.03) & 4810 &  3.12 & --0.07 \\
HD~142373    &  15.89 \scriptsize{(0.02)} & 3.90 (0.02) & 5830 &  3.96 & --0.54 \\
HD~218857    &  346.2 \scriptsize{(2.3) } &  2.57 (0.03) & 5060 &  2.53 & --1.92 \\
HE0011-0035  &   7080 \scriptsize{(1308)} &  2.32 (0.16) & 4950 &  2.00 & --3.04 \\
HE0039-4154  &   7032 \scriptsize{(720) } &  1.80 (0.09) & 4780 &  1.60 & --3.26 \\
HE0048-0611  &   6434 \scriptsize{(1253)} &  2.69 (0.17) & 5180 &  2.70 & --2.69 \\
HE0122-1616  &   5582 \scriptsize{(1103)} &  2.94 (0.17) & 5200 &  2.65 & --2.85 \\
HE0332-1007  &  12774 \scriptsize{(2892)} &  1.47 (0.20) & 4750 &  1.50 & --2.89 \\
HE0445-2339  &   6673 \scriptsize{(624) } &  2.03 (0.09) & 5165 &  2.20 & --2.76 \\
HE1356-0622  &   7770 \scriptsize{(1507)} &  1.88 (0.17) & 4945 &  2.00 & --3.45 \\
\hline \noalign{\smallskip}             
\end{tabular}                           

{\bf Notes.} The errors of the Gaia eDR3 distances and $\logg_{\rm d}$ are indicated in parentheses.
\end{table}

\begin{table} 
 \centering
 \contcaption{\label{tab:continued} -- Distance-based and spectroscopic surface gravities for the Galactic stellar samples from \citet{2015ApJ...808..148S} and \citet{dsph_parameters}.}
 \begin{tabular}{lccccr}
\hline\hline \noalign{\smallskip}
 \multicolumn{1}{c}{Star} & \multicolumn{2}{c}{Gaia eDR3} &  $\Teff$ & $\logg$, & [Fe/H] \\
\cline{2-3} 
             &   d(pc)  & $\logg_{\rm d}$ &  (K)  &     Sp     &        \\
\noalign{\smallskip} \hline 
\multicolumn{1}{c}{1} & \multicolumn{1}{c}{2} & 3 &  4      &  5   & \multicolumn{1}{c}{6} \\      
 \hline \noalign{\smallskip}
HE1357-0123         &  14890 \scriptsize{(4225)} &  1.30 (0.25) & 4600 &  1.20 & --3.92 \\
HE1416-1032         &   8244 \scriptsize{(1635)} &  2.15 (0.17) & 5000 &  2.00 & --3.23 \\
HE2244-2116         &   7703 \scriptsize{(2213)} &  2.67 (0.25) & 5230 &  2.80 & --2.40 \\
HE2249-1704         &  10602 \scriptsize{(2771)} &  1.86 (0.23) & 4590 &  1.20 & --2.94 \\
HE2252-4225         &   9265 \scriptsize{(2354)} &  1.93 (0.22) & 4750 &  1.55 & --2.76 \\
HE2327-5642         &   4711 \scriptsize{(308) } &  2.25 (0.06) & 5050 &  2.20 & --2.92 \\
BD+07$^\circ$4841   &  156.0 \scriptsize{(0.4) } &  4.30 (0.02) & 6130 &  4.15 & --1.46 \\
BD+09$^\circ$0352   &  158.1 \scriptsize{(0.5) } &  4.16 (0.02) & 6150 &  4.25 & --2.09 \\
BD+24$^\circ$1676   &  250.2 \scriptsize{(1.1) } &  4.08 (0.02) & 6210 &  3.90 & --2.44 \\
BD+29$^\circ$2091   &  87.69 \scriptsize{(0.15)} &  4.59 (0.02) & 5860 &  4.67 & --1.91 \\
BD+37$^\circ$1458   &  143.9 \scriptsize{(0.3) } &  3.56 (0.03) & 5500 &  3.70 & --1.95 \\
BD+66$^\circ$0268   &  49.55 \scriptsize{(0.04)} &  4.66 (0.03) & 5300 &  4.72 & --2.06 \\
BD$-04^\circ$3208   &  174.1 \scriptsize{(0.6) } &  4.12 (0.02) & 6390 &  4.08 & --2.20 \\
BD$-11^\circ$0145   &   2055 \scriptsize{( 80) } &  1.66 (0.04) & 4900 &  1.73 & --2.18 \\
BD$-13^\circ$3442   &  210.6 \scriptsize{(1.4) } &  4.12 (0.02) & 6400 &  3.95 & --2.62 \\
CD$-24^\circ$1782   &  469.0 \scriptsize{(3.7) } &  2.76 (0.03) & 5140 &  2.62 & --2.72 \\
BS16550-0087        &   8253 \scriptsize{(1052)} &  1.58 (0.11) & 4750 &  1.50 & --3.33 \\
G090-003            &  250.8 \scriptsize{(1.1) } &  3.89 (0.02) & 6007 &  3.90 & --2.04 \\
\hline \noalign{\smallskip}
\end{tabular}

{\bf Notes.} The errors of the Gaia eDR3 distances and $\logg_{\rm d}$ are indicated in parentheses.
\end{table}

\begin{figure}  
 \begin{minipage}{85mm}
\centering
	\includegraphics[width=0.99\textwidth, clip]{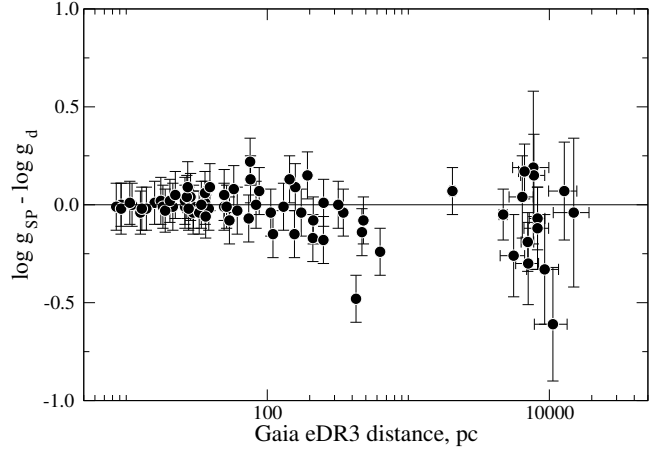}
  \caption{\label{Fig:gaia} The differences between spectroscopic and distance-based surface gravities depending on the Gaia eDR3 distance for the Galactic stellar samples from \citet{2015ApJ...808..148S} and \citet{dsph_parameters}. }
\end{minipage}
\end{figure}

\subsection{Spectroscopic versus Gaia eDR3 distances}\label{sect:distances}

Iron is represented in the VMP stars by the two ionization stages, which are used in many studies to determine  spectroscopic surface gravities ($g_{\rm Sp}$) from the requirement that abundances from lines of Fe\ione\ and Fe\ii\ in a given star must be equal. The surface gravity can also be derived from distance; this is the distance-based surface gravity, $g_{\rm d}$. If $g_{\rm Sp}$ based on the NLTE calculations and $g_{\rm d}$ are obtained to be consistent within the error bars, this means that the calculations for Fe\ione -Fe\ii\ are correct.

\citet{2015ApJ...808..148S} and \citet{dsph_parameters} derived the surface gravities for the two Galactic stellar samples using photometric effective temperatures and the NLTE analysis of the Fe\ione\ and Fe\ii\ lines. Using the Gaia eDR3 parallaxes corrected according to
\citet{2021A&A...649A...4L}, we calculated distances from the maximum
of the distance probability distribution function, as recommended by
\citet{2015PASP..127..994B}, and then
$\logg_{\rm d}$ from the relation

$$
	\mathrm{log}\,g_{\rm d} = -10.607 +\mathrm{log}\,M+4\,\mathrm{log}\,T_{\rm eff} -
$$
$$
	0.4\,[4.74 - (V + BC + 5 - 5\, \mathrm{log\, d} - A_V)]
$$

\noindent Here, $M$ is a star's mass, $A_V$ is an interstellar extintion in the V-band, $BC$ is a bolometric correction which was calculated by interpolation in the grid of \citet{1998A&A...333..231B}\footnote{\url{https://wwwuser.oats.inaf.it/castelli/colors/bcp.html}}. The atmospheric parameters and $A_V$ were taken from \citet{2015ApJ...808..148S} and \citet{dsph_parameters}. Stellar masses and $V$ magnitudes for the \citet{2015ApJ...808..148S} sample are listed in their Table~5 and 2, respectively. For the stellar sample of \citet{dsph_parameters}, the $V$ magnitudes are listed in their Table~5. For each VMP giant, we adopt $M = 0.8 M_{\odot}$.

Statistical error of the distance-based surface gravity was computed as the quadratic sum of errors of the star's distance, effective temperature, mass, visual magnitude, and $BC$. We assumed the stellar mass error as $\sigma_M = 0.1 M_{\odot}$ and took the effective temperature errors, $\sigma_T$, from \citet{2015ApJ...808..148S} and \citet{dsph_parameters}. The total error is dominated by $\sigma_M$ for the nearby stars and by the distance error, $\sigma_{\rm d}$, for the distant objects.

Table~\ref{Tab:gaia} lists the obtained Gaia eDR3 distances and $\logg_{\rm d}$ values, as well as the spectroscopic surface gravities from \citet{2015ApJ...808..148S} and \citet{dsph_parameters}.
The differences log~$g_{\rm Sp}$ -- $\logg_{\rm d}$ are shown in Fig.~\ref{Fig:gaia}. The majority of our stars lie within 631~pc from the Sun, and their spectroscopic surface gravities are found to be consistent within the error bars with the distance-based ones. A clear outlier is HD~8724, with log~$g_{\rm Sp}$ -- $\logg_{\rm d} = -0.48$. We note that the discrepancy between log~$g_{\rm Sp}$ and $\logg_{\rm d}$ has reduced compared to $-0.76$~dex obtained for HD~8724 by \citet{dsph_parameters} using the Gaia DR1 parallax \citep{2016A&A...595A...2G}. However, it is still greater than the error of spectroscopic surface gravity, $\sigma_{\log g{\rm (sp)}}$ = 0.24~dex. Formal calculation of $\sigma_{\log g{\rm (d)}}$ leads to 0.07~dex (Table~\ref{Tab:gaia}), however, astrometric\_excess\_noise\_sig = 6.005 and astrometric\_chi2\_al = 419.84 indicated by \citet{2021A&A...649A...4L} for HD~8724 suggest an unreliable solution for the Gaia eDR3 parallax.

For 15 distant stars, with d $>$ 2 kpc, the errors of $\logg_{\rm d}$ grow. Nevertheless, the spectroscopic surface gravities are consistent, on average, with the distance-based ones.

Thus, our NLTE method for Fe\ione /Fe\ii\ is reliable and can be used for determinations of surface gravities, in particular, for distant stars with large distance errors.

\begin{figure}  
 \begin{minipage}{85mm}
\centering
	\includegraphics[width=0.99\textwidth, clip]{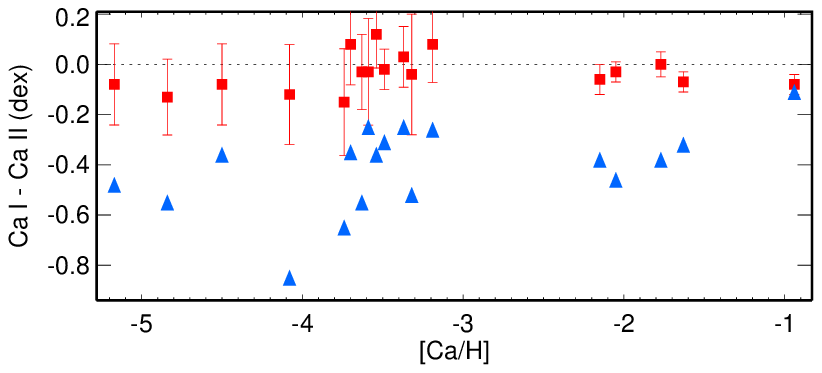}
  \caption{\label{Fig:ca1_2} NLTE (red squares) and LTE (blue triangles) differences between the Ca\ione - and Ca\ii -based abundances for the stellar samples from \citet{2017AA...605A..53M} and \citet{2019MNRAS.485.3527S}. }
\end{minipage}
\end{figure}

\subsection{Ca\ione\ versus Ca\ii}

A firm argument for a correct treatment of the NLTE line formation for Ca\ione -Ca\ii\ can be obtained from a comparison of the NLTE abundances from lines of the two ionization stages. \citet{2017AA...605A..53M} report the LTE and NLTE abundances from lines of Ca\ione\ and Ca\ii\ 8498~\AA\ for five reference stars with well-determined atmospheric parameters in the $-2.7 <$ [Fe/H] $< -1.3$ metallicity range and find fairly consistent NLTE abundances, while the LTE abundance difference between Ca\ione\ and Ca\ii\ 8498~\AA\ grows in absolute value towards lower metallicity and reaches $-0.45$~dex for [Fe/H] = $-2.62$, see their Fig.~6. 

\citet{2019MNRAS.485.3527S} studied the UMP stars and improved their atmospheric parameters using an extensive method based on the colour-$\Teff$ calibrations, NLTE fits of the Balmer line wings, and Gaia DR2 trigonometric parallaxes. For each star, the derived effective temperature and surface gravity were checked by inspecting the Ca\ione /Ca\ii\ NLTE ionization equilibrium and by comparing the star's position in the $\logg - \Teff$ plane
with the theoretical isochrones of 12 and 13~Gyr.  

The abundance differences between the two ionization stages from the NLTE and LTE calculations of \citet{2017AA...605A..53M} and \citet{2019MNRAS.485.3527S} are displayed in Fig.~\ref{Fig:ca1_2}. Nowhere, the NLTE abundance difference Ca\ione\ -- Ca\ii\ exceeds 0.15~dex, while the LTE abundances from lines of Ca\ii\ are systematically lower compared with that from Ca\ione, by up to 0.85~dex. Thus, the NLTE results obtained using our NLTE method for Ca\ione -\ii\ \citep{2017AA...605A..53M} can be trusted.

\begin{figure*}
 \begin{minipage}{150mm}

\hspace{-10mm}
\parbox{0.5\linewidth}{\includegraphics[scale=0.5]{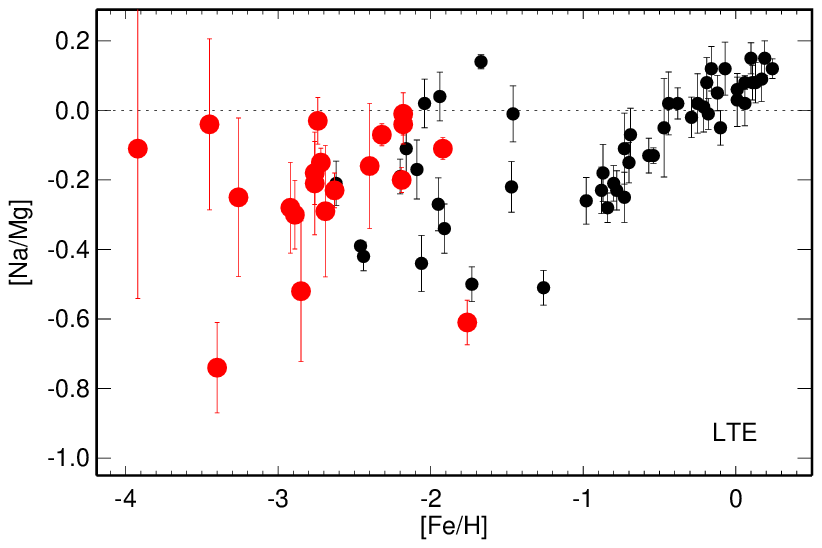}\\
\centering}
\hspace{5mm}
\parbox{0.5\linewidth}{\includegraphics[scale=0.5]{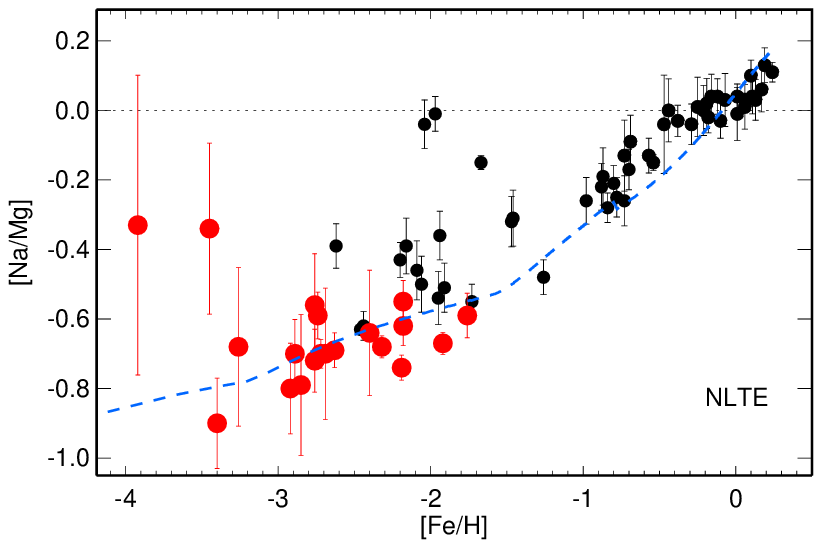}\\
\centering}

\vspace{-7mm}
  \caption{\label{Fig:namg} [Na/Mg] LTE (left panel) and NLTE (right panel) abundance ratios for the dwarf stellar sample from \citet[][small black circles]{lick_paperII} and the giant sample from \citet[][large red circles]{2017A&A...608A..89M}. The NLTE abundance ratios are well reproduced by the chemical evolution model of \citet[][dashed curve]{2020ApJ...900..179K}. }
\end{minipage}
\end{figure*}

\subsection{Na\ione\ resonance lines in VMP stars}

Figure~\ref{Fig:namg} displays the [Na/Mg] abundance ratios in the wide range of metallicities from the LTE and NLTE calculations of \citet{lick_paperII} and \citet{2017A&A...608A..89M}. For [Fe/H] $> -1$, both LTE and NLTE data form a well-defined upward trend, with a small star-to-star scatter for the stars of close metallicity. The situation is very different in LTE and NLTE for [Fe/H] $< -1$. In LTE, the [Na/Mg] ratios reveal a big scatter, which is substantially reduced in the NLTE calculations. An explanation lies mostly with the NLTE effects for lines of Na\ione. For Mg, the differences between the NLTE and LTE abundances do not exceed 0.1~dex.

For [Fe/H] $> -1$, the Na abundances were derived by \citet{lick_paperII} from the Na\ione\ 5682, 5688, 6154, 6160 5895~\AA\ subordinate lines, which are slightly affected by NLTE, with negative $\Delta_{\rm NLTE}$ of $\precsim$0.1~dex, in absolute value.
In the lower metallicity stars, sodium is observed in the Na\ione\ 5889, 5895~\AA\ resonance lines only. They are subject to strong NLTE effects, with $\Delta_{\rm NLTE}$ depending on the atmospheric parameters and the Na abundance itself. For different stars, $\Delta_{\rm NLTE}$ varies between $-0.1$ and $-0.6$~dex \citep{2017A&A...608A..89M}. Removing the star-to-star scatter of the [Na/Mg] NLTE abundance ratios for [Fe/H] $< -1$ can serve as a circumstantial evidence for the line formation to be treated correctly.

Taking advantage of the obtained Galactic NLTE [Na/Mg] trend, we found that the modern nuclesynthesis and Galactic chemical evolution (GCE) calculations, which are represented in Fig.~\ref{Fig:namg} (right panel) by the GCE model of \citet{2020ApJ...900..179K}, predict correctly contributions from the core-collapse supernovae (SNeII) and the asymptotic giant branch (AGB) stars to production of Mg and Na during the Galaxy history.

\begin{figure}  
 \begin{minipage}{85mm}
\centering
	\includegraphics[width=0.99\textwidth, clip]{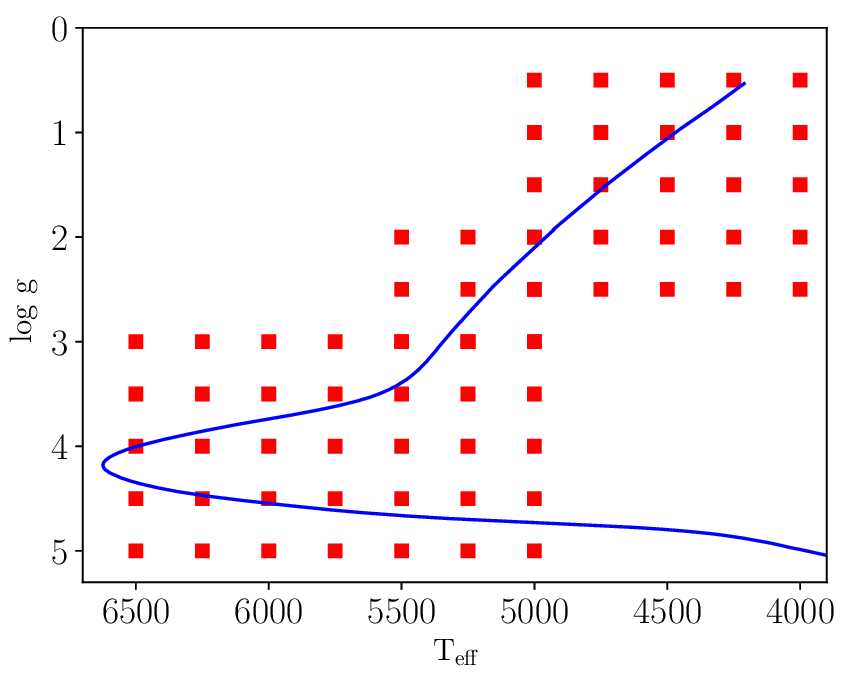}
  \caption{\label{Fig:isochrone} Stellar parameter range covered by our tables of the NLTE abundance corrections is shown by squares. The solid curve is the isochrone of 12~Gyr, [Fe/H] = $-2$, and [$\alpha$/Fe] = 0.4 from \citet{2008ApJS..178...89D}. }
\end{minipage}
\end{figure}

\section{Grids of the NLTE abundance corrections}\label{Sect:grids}

By request of the Pristine collaboration \citep{2017MNRAS.471.2587S}, the NLTE abundance corrections were computed for the lines which can be detected in spectra of VMP stars, that is, for the [Fe/H] $\le -2$ range. We focused, in particular, on the spectral ranges observed by WEAVE\footnote{https://ingconfluence.ing.iac.es/confluence/display/WEAV/Science}, that is 4040-4650~\AA, 4750-5450~\AA, and 5950-6850~\AA\ for the high-resolution ($R = \lambda/\Delta \lambda$ = 20\,000) observations and 3660-9590~\AA\ for the $R$ = 5000 observations, and 4MOST \footnote{https://www.4most.eu/cms}, that is 3926-4350~\AA, 5160-5730~\AA, and 6100-6790~\AA\ for the high-resolution spectrograph (HRS, $R \simeq$ 20\,000) and 3700-9500~\AA\ for the low-resolution spectrograph (LRS, $R \simeq$ 4000-7500). We selected
 4 / 15 / 28 / 4 / 54 / 262 / 7 / 2 / 2 / 5 lines of Na\ione\ / Mg\ione\ / Ca\ione\ / Ca\ii\ / Ti\ii\ / Fe\ione\ / Zn\ione\ / Zn\ii\ / Sr\ii\ / Ba\ii.

The range of atmospheric parameters was selected to represent metal-poor stars on various evolutionary stages, from the main sequence to the red giant branch (RGB); see the isochrone of 12~Gyr, [Fe/H] = $-2$, and [$\alpha$/Fe] = 0.4 from \citet{2008ApJS..178...89D} in Fig.~\ref{Fig:isochrone}. The NLTE calculations were performed in the following ranges of effective temperature and surface gravity:

$\Teff$ = 4000 to 4750~K for $\logg$ = 0.5 to 2.5;

$\Teff$ = 5000~K for $\logg$ = 0.5 to 5.0;

 $\Teff$ = 5250 to 5500~K for $\logg$ = 2.0 to 5.0;

 $\Teff$ = 5750 to 6500~K for $\logg$ = 3.0 to 5.0.

 Metallicity range is $-5.0 \le$ [Fe/H] $\le -2.0$.


The nodes of the NLTE abundance correction grids correspond to the nodes of the MARCS model grid. Therefore, $\Teff$ varies with a step of 250~K, $\logg$ with a step of 0.5, and [Fe/H] with a step of 0.5. The MARCS website does not provide models with [Fe/H] = $-3.5$ and $-4.5$. The missing models were calculated by interpolating between the [Fe/H] = $-3$ and $-4$ and between the [Fe/H] = $-4$ and $-5$ models. We applied the FORTRAN-based interpolation routine written by Thomas Masseron and available on the MARCS website.

For Fe\ione -\ii\ and Zn\ione -\ii, the SE calculations were performed with [Element/Fe] = 0.0;
for Mg\ione\ and Ti\ii\ with [Element/Fe] = 0.4 and 0.3, respectively.

For Na\ione, Ca\ione, Ca\ii, Sr\ii, and Ba\ii, the NLTE effects are sensitive to not only $\Teff$/$\logg$/[Fe/H], but also the element abundance used in the SE calculations. Therefore, the grids of the NLTE corrections are 4-dimensional where [Element/Fe] takes the following numbers:

\noindent
[Na/Fe] = $-0.6$, $-0.3$, 0.0, 0.3, 0.6;

\noindent
[Ca/Fe] = 0.0 and 0.4;

\noindent
[Sr/Fe] = $-1.0$, $-0.5$, 0.0, 0.5, 1.0 for the dwarf model atmospheres,

\noindent [Sr/Fe] = $-1.5$, $-1.0$, $-0.5$, 0.0, 0.5 for the giant model atmospheres;

\noindent
[Ba/Fe] =  $-1.0$, $-0.5$, 0.0, 0.5 for the dwarf model atmospheres,

\noindent [Ba/Fe] = $-1.5$, $-1.0$, $-0.5$, 0.0, 0.5 for the giant model atmospheres.

The website INASAN\_NLTE\footnote{\url{http://spectrum.inasan.ru/nLTE2/}} provides the tools for calculating online the NLTE abundance correction(s) for given spectral line(s) and atmospheric parameters $\Teff$, $\logg$, [Fe/H], [Element/Fe] by an interpolation in the NLTE correction grids.

\begin{figure}  
 \begin{minipage}{85mm}
\centering
	\includegraphics[width=0.99\textwidth, clip]{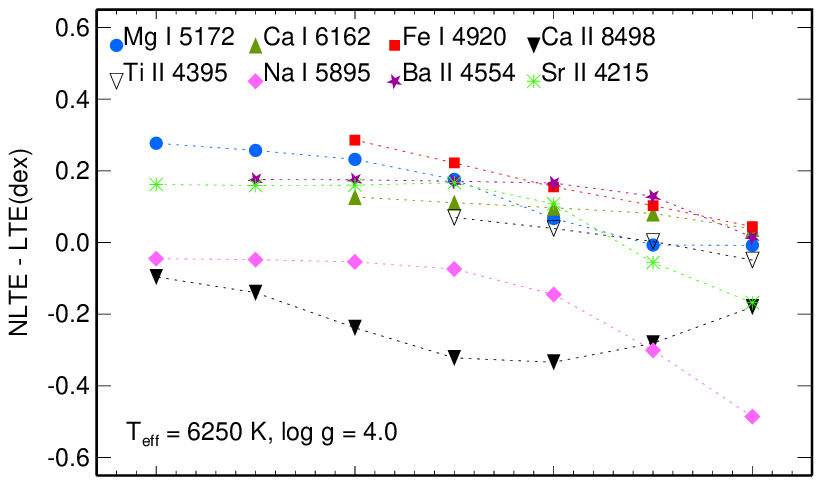}

  \vspace{-12mm}
	\includegraphics[width=0.99\textwidth, clip]{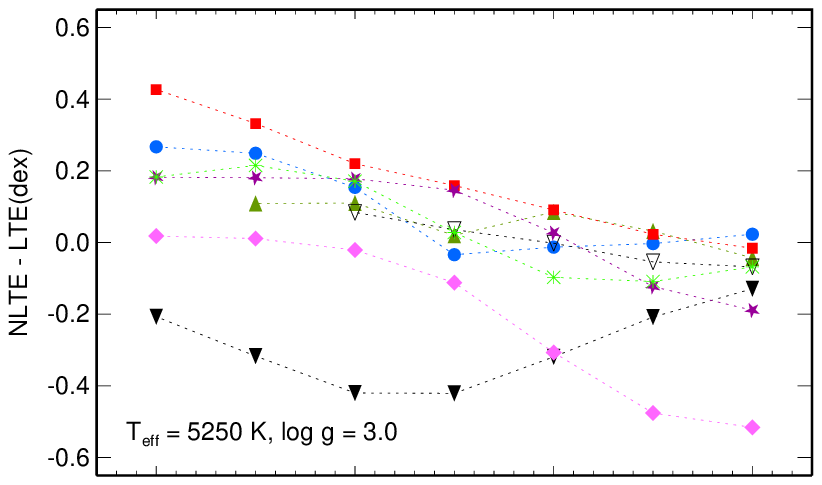}

  \vspace{-12mm}
	\includegraphics[width=0.99\textwidth, clip]{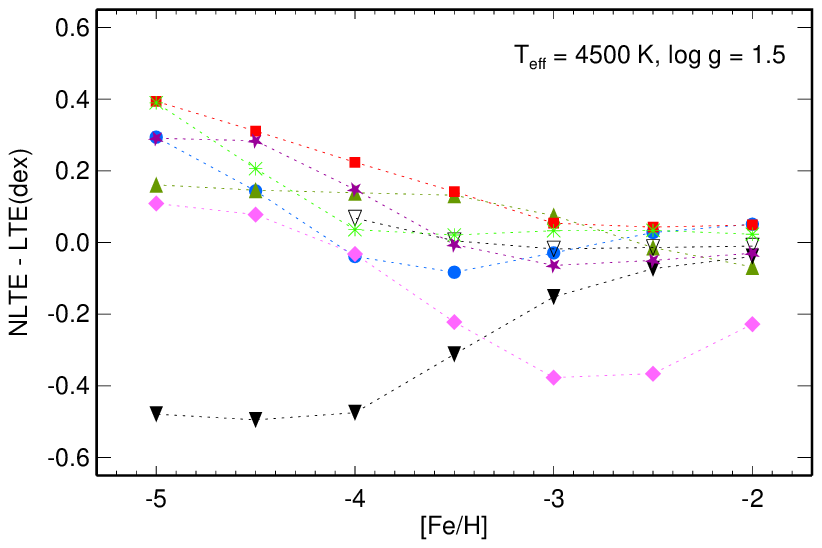}

  \caption{\label{Fig:teff} NLTE abundance corrections for lines of Na\ione\ (rhombi), Mg\ione\ (circles), Ca\ione\ (triangles), Ca\ii\ (filled inverted triangles), Ti\ii\ (open inverted triangles), Fe\ione\ (squares), Sr\ii\ (asterisks), and Ba\ii\ (five-pointed star) as a function of metallicity in the models with $\Teff$/$\logg$ = 6250/4.0 (top panel), 5250/3.0 (middle panel), and 4500/1.5 (bottom panel). Everywhere in the NLTE calculations, [Element/Fe] = 0 was adopted. }
\end{minipage}
\end{figure}

\subsection{NLTE corrections depending on atmospheric parameters}\label{Sect:feh}

Figure~\ref{Fig:teff} displays the NLTE abundance corrections predicted for representative lines of different chemical species in VMP stars on different evolutionary stages, namely, the turn-off (TO, $\Teff$/$\logg$ = 6250/4.0), the bottom red giant branch (bRGB, 5250/3.0), and the RGB (4500/1.5). For each line, $\Delta_{\rm NLTE}$ depends on $\Teff$, $\logg$ and [Fe/H]. Therefore, neglecting the NLTE effects distorts the galactic abundance trends.
In the same atmosphere, different lines have the NLTE corrections of different magnitude and sign. Therefore,
the star's element abundance pattern derived under the LTE assumption does not reflect correctly relative contributions of different nuclesynthesis sources.

The sign of $\Delta_{\rm NLTE}$ is determined by the mechanisms that produce the departures from LTE for lines of a given species in given physical conditions. 

In the stellar parameter range with which we concern, Mg\ione, Ca\ione, and Fe\ione\ are the minority species in the line formation layers, and they are subject to the ultra-violet (UV) overionization, resulting in depleted atomic level populations, weakened lines, and positive NLTE abundance corrections \citep[see][for detailed analyses]{1996ARep...40..187M,mash_ca,mash_fe}. The intensity of the ionizing UV radiation increases with decreasing metallicity, resulting in growing departures from LTE. 

Na\ione\ is also the minority species, however, due to low photoionization cross-sections of its ground state, the main NLTE mechanism is a "photon suction" process \citep{1992A&A...265..237B} which produces overpopulation of the neutral stage, resulting in strengthened Na\ione\ lines and negative NLTE abundance corrections. Photon suction is connected with collisional processes that couple the high-excitation levels of Na\ione\ with the singly ionized stage. In contrast to the radiative processes, an influence of collisional processes on the statistical equilibrium of Na\ione\ is weakened with decreasing metallicity, and $\Delta_{\rm NLTE}$ for Na\ione\ 5895~\AA\ decreases in absolute value and becomes even slightly positive for [Fe/H] $\le -4.5$ in the 4500/1.5 models.

The NLTE effects for the majority species Ca\ii, Ti\ii, Sr\ii, and Ba\ii\ are driven by the bound-bound (b-b) transitions. For an individual line, the sign and magnitude of $\Delta_{\rm NLTE}$ depend on the physical conditions and the transition where the line arises. Ca\ii\ 8498~\AA\ arises in the transition \eu{3d}{2}{D}{}{3/2} -- \eu{4p}{2}{P}{\circ}{3/2}. The upper level is depopulated in the atmospheric layers where the core of  Ca\ii\ 8498~\AA\ forms via photon loss in the wings of the Ca\ii\ 3933, 3968~\AA\ resonance lines and the 8498, 8542, 8668~\AA\ infra-red (IR) triplet lines. The Ca\ii\ 8498~\AA\ line core is strengthened because the line
source function drops below the Planck function, resulting in negative $\Delta_{\rm NLTE}$ \citep{mash_ca}. In the [Fe/H] = $-2$ models, Ca\ii\ 8498~\AA\ is very strong with a total absorption dominated by the line wings that form in deep atmospheric layers where the NLTE effects are small. With decreasing [Fe/H] (and Ca abundance, too) the line wings are weakened, and $\Delta_{\rm NLTE}$ grows in absolute value. In the 6250/4.0 and 5250/3.0 models, $\Delta_{\rm NLTE}$ decreases in absolute value for [Fe/H] $\le -3.5$ because of shifting the formation depths for Ca\ii\ 8498~\AA\ in deep atmospheric layers. 

Owing to a complex atomic term structure, the levels of Ti\ii\ are tightly coupled to each other and to the ground state via radiative and collisional processes, and the NLTE corrections for the Ti\ii\ lines are slightly positive in the stellar parameter range with which we concern \citep{sitnova_ti}: $\Delta_{\rm NLTE} \precsim$ 0.1~dex for Ti\ii\ 4395~\AA.

\citet{Mashonkina2001sr} and \citet{Mashonkina1999} predicted theoretically that NLTE may either strengthen or weaken the lines of Sr\ii\ and Ba\ii, depending on the stellar parameters and elemental abundance. For example, in the 6250/4.0 models, $\Delta_{\rm NLTE}$ is positive for Ba\ii\ 4554~\AA\ over full range of [Fe/H] = $-2$ down to $-4.5$, while, for Sr\ii\ 4215~\AA, $\Delta_{\rm NLTE}$ is negative when [Fe/H] $\ge -2.5$ and positive for the more metal-deficient atmospheres. In the RGB atmospheres, both Sr\ii\ 4215~\AA\ and Ba\ii\ 4554~\AA\ are very strong until metallicity decreases to [Fe/H] = $-3.5$, and the NLTE corrections are small. For the lower metallicity, $\Delta_{\rm NLTE}$ is positive for both lines and grows with decreasing [Fe/H].

For lines of Zn\ione, the NLTE abundance corrections depending on atmospheric parameters are discussed by \citet{2022MNRAS.515.1510S}.

\begin{figure*}  
 \begin{minipage}{150mm}

\hspace{-10mm}
\parbox{0.3\linewidth}{\includegraphics[scale=0.37]{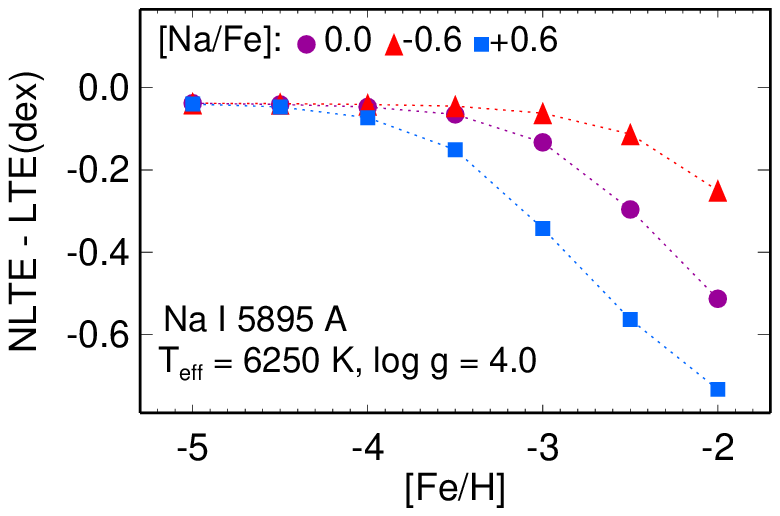}\\
\centering}
\hspace{5mm}
\parbox{0.3\linewidth}{\includegraphics[scale=0.37]{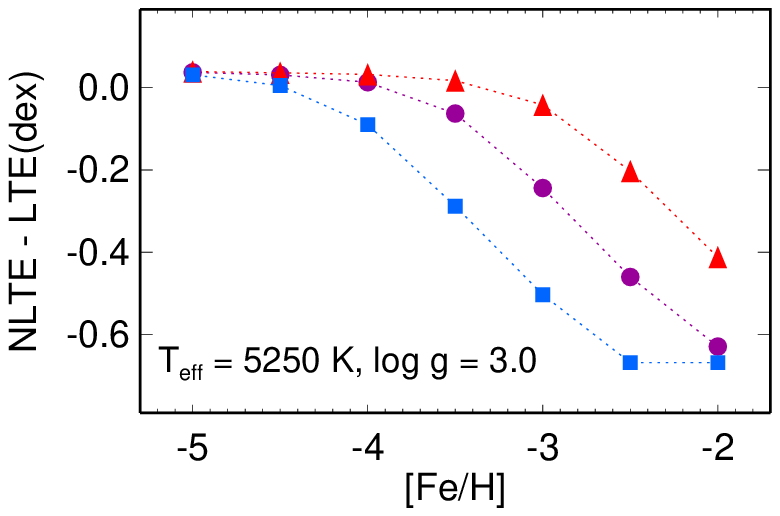}\\
\centering}
\hspace{5mm}
\parbox{0.3\linewidth}{\includegraphics[scale=0.37]{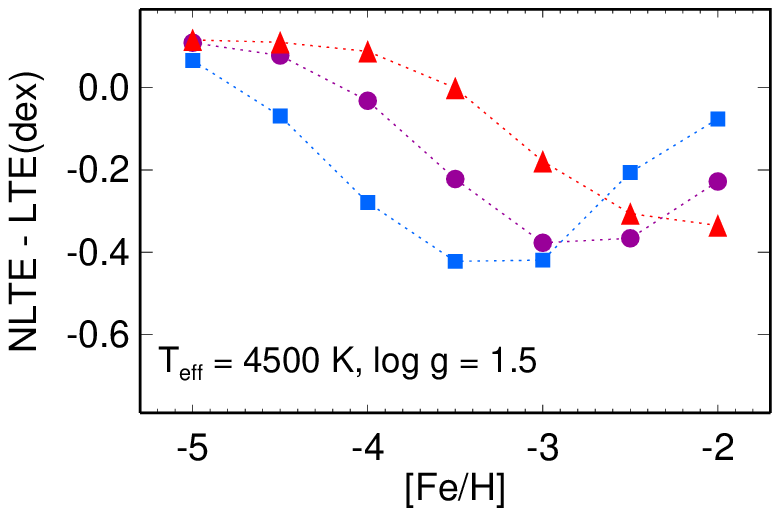}\\
\centering}
\hfill
\\[0ex]

\vspace{-7mm}
  \caption{\label{Fig:na5895} NLTE abundance corrections for Na\ione\ 5895~\AA\ depending on the element abundance in the models representing atmospheres of the TO, bRGB, and RGB stars. The circles, triangles, and squares correspond to [Na/Fe] = 0.0, $-0.6$, and +0.6, respectively. }
\end{minipage}
\end{figure*}

\begin{figure*}           
 \begin{minipage}{150mm}

\hspace{-10mm}
\parbox{0.3\linewidth}{\includegraphics[scale=0.37]{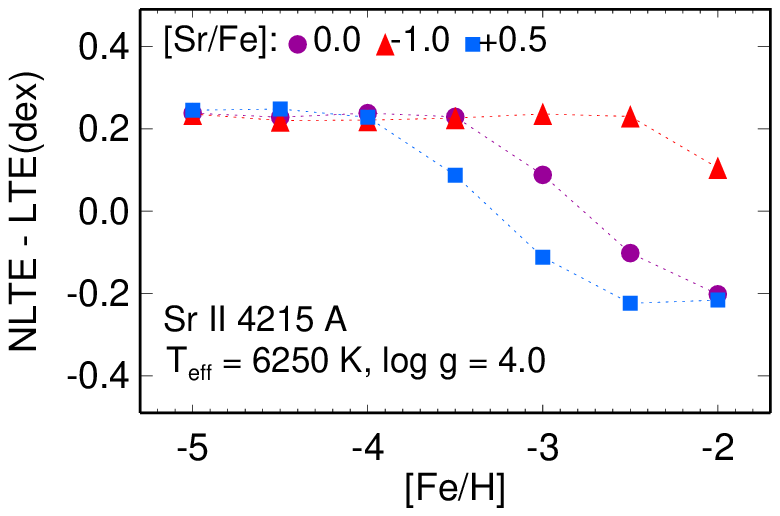}\\
\centering}
\hspace{5mm}
\parbox{0.3\linewidth}{\includegraphics[scale=0.37]{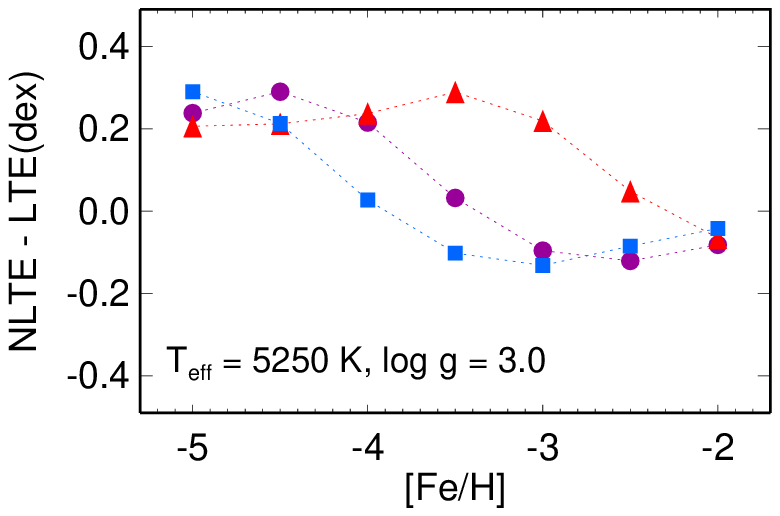}\\
\centering}
\hspace{5mm}
\parbox{0.3\linewidth}{\includegraphics[scale=0.37]{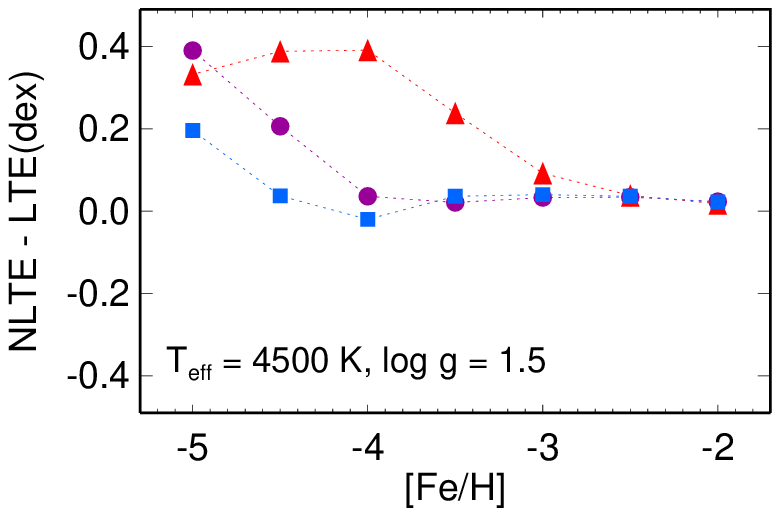}\\
\centering}

\vspace{-7mm}

\hspace{-10mm}
\parbox{0.3\linewidth}{\includegraphics[scale=0.37]{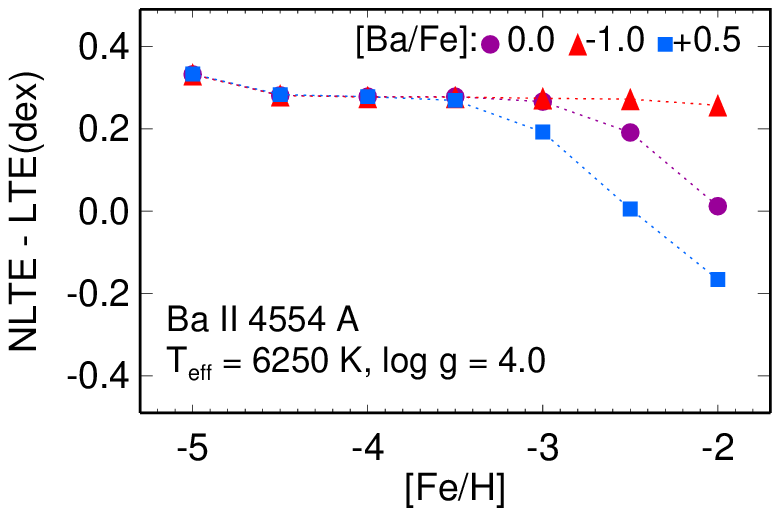}\\
\centering}
\hspace{5mm}
\parbox{0.3\linewidth}{\includegraphics[scale=0.37]{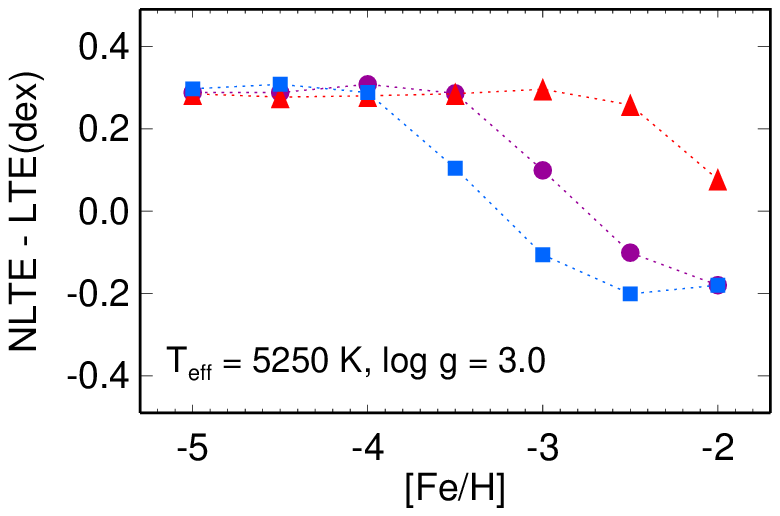}\\
\centering}
\hspace{5mm}
\parbox{0.3\linewidth}{\includegraphics[scale=0.37]{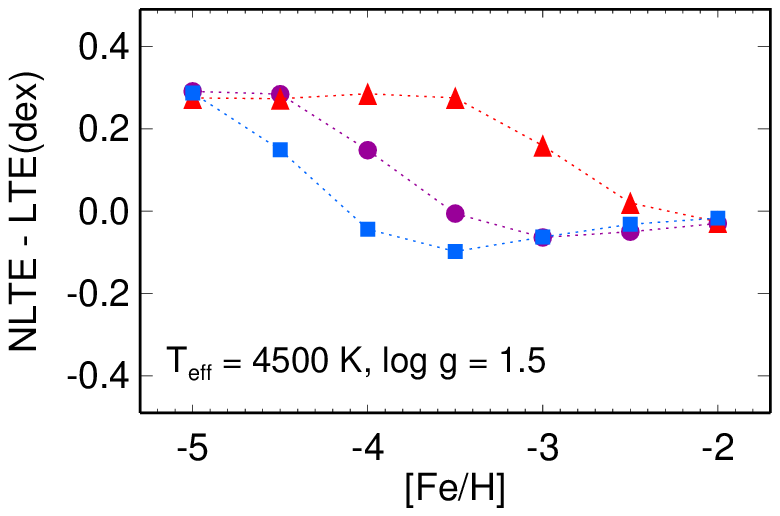}\\
\centering}

\hfill
\\[0ex]

\vspace{-7mm}
  \caption{\label{Fig:ba4554} The same as in Fig.~\ref{Fig:na5895} for Sr\ii\ 4215~\AA\ (top row) and Ba\ii\ 4554~\AA\ (bottom row). The circles, triangles, and squares correspond to [Element/Fe] = 0.0, $-1.0$, and +0.5, respectively. }
\end{minipage}
\end{figure*}

\subsection{NLTE corrections depending on elemental abundances}\label{Sect:xh}

The stars of close metallicity in the [Fe/H] $< -2$ range reveal a substantial scatter of the Na, Sr, and Ba abundances \citep[see, for example][]{Cohen2013}. Exactly for Na\ione, Sr\ii, and Ba\ii\ the NLTE effects depend strongly on not only atmospheric parameters, but also the element abundance. Therefore in order to interpret correctly the chemical evolution of Na, Sr, and Ba, abundance analyses of VMP samples should be based on the NLTE abundances. 

Figure~\ref{Fig:na5895} shows that, for the TO and bRGB stars, the LTE analysis overestimates the Na abundances, by the quantity which is greater for the Na-enhanced than for Na-poor star. The difference in $\Delta_{\rm NLTE}$ exceeds 0.4~dex for [Fe/H] = $-2.5$ and reduces towards the lower [Fe/H]. The same is true for the RGB stars with [Fe/H] $\le -3.5$, but the situation is more complicated for the higher metallicities. For [Fe/H] $> -3$, the Na\ione\ 5895~\AA\ line is very strong in the Na-enhanced cool atmospheres, and the total line absorption is dominated by the line wings that form in deep atmospheric layers affected only weakly by NLTE. Accounting for the NLTE effects for the Na\ione\ lines reduces substantially the abundance discrepancies found for stellar samples in LTE, as well illustrated by Fig.~\ref{Fig:namg}.

Using the same atmospheric parameters, LTE may either overestimate, or underestimate abundances of Sr and Ba depending on the elemental abundances, as shown in Fig.~\ref{Fig:ba4554}. For [Fe/H] $< -2$, the NLTE abundance corrections for Sr\ii\ 4215~\AA\ and Ba\ii\ 4554~\AA\ are positive in the Sr- and Ba-poor atmospheres, while they can be negative for the Sr- and Ba-enhanced atmospheres. Accounting for the NLTE effects can reduce the abundance discrepancies found for stellar samples in LTE, by more than 0.4~dex for Sr in the TO [Fe/H] = $-2.5$ stars and for Ba in the bRGB [Fe/H] = $-2.5$ stars. 

\begin{figure*}  
 \begin{minipage}{150mm}
\centering
	\includegraphics[scale=0.43]{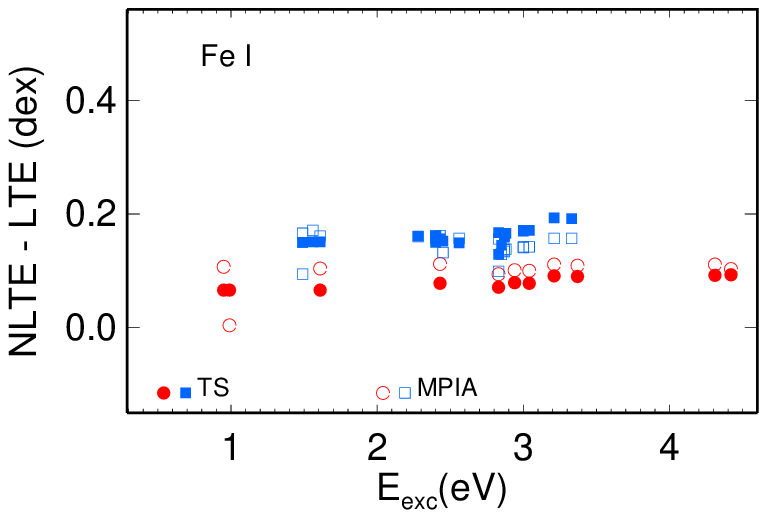}
	\includegraphics[scale=0.43]{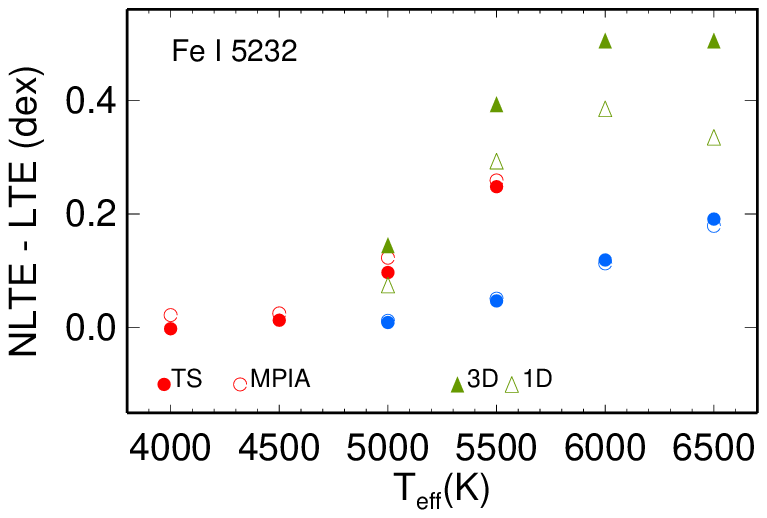}
  \caption{\label{Fig:compare1} NLTE abundance corrections for lines of Fe\ione\ from our calculations (this study = TS, filled circles and squares), the NLTE\_MPIA database (open circles and squares), and \citet[][triangles]{2022A&A...668A..68A}. Left panel: lines of various \Eexc\ in the model atmospheres 6350/4.09/$-2.18$ (circles) and 4630/1.28/$-2.99$ (squares). Right panel: Fe\ione\ 5232~\AA\ in the model atmospheres with [Fe/H] = $-3$ for $\logg$ = 2.5 (red symbols) and 4.0 (blue and green symbols). The filled and open triangles corresponds to the 3D NLTE and 1D NLTE calculations, respectively.}
\end{minipage}
\end{figure*}

\subsection{NLTE corrections for different type model atmospheres}

The model atmospheres computed with different codes produce, as a rule, very similar atmospheric structures and spectral energy distributions for common atmospheric parameters. We checked how different type model atmospheres influence on magnitudes of the NLTE abundance corrections. Taking the ATLAS9-ODFNEW models from R. Kurucz's website\footnote{\url{http://kurucz.harvard.edu/grids/gridm40aodfnew/}}, we performed the NLTE calculations for Ca\ione -\ii, Fe\ione -\ii, and Ba\ii\ with the models 6250/4.0/$-4.0$ and 4500/1.5/$-4.0$. For these atmospheric parameters, the selected lines reveal the greatest NLTE effects. The results are presented in Table~\ref{Tab:kurucz}.

\begin{table} 
 \caption{\label{Tab:kurucz} NLTE abundance corrections for the selected lines in the MARCS and ATLAS9-ODFNEW model atmospheres.}
 \begin{tabular}{lrrcrr}\hline\hline \noalign{\smallskip}
 Line & \multicolumn{2}{c}{6250/4.0/$-4.0$} & & \multicolumn{2}{c}{4500/1.5/$-4.0$} \\
\cline{2-3}                
\cline{5-6}                
                  &  MARCS & ATLAS9 & & MARCS & ATLAS9 \\
\noalign{\smallskip} \hline \noalign{\smallskip}
Ca\ione\ 4226~\AA &  0.163 &  0.199 & &  0.039 & $-$0.019 \\ 
Ca\ii\ 8498~\AA   & $-$0.238 & $-$0.220 & & $-$0.475 & $-$0.531 \\ 
Fe\ione\ 4920~\AA &  0.286 &  0.322 & &  0.224 &  0.229 \\
Ba\ii\ 4554~\AA   &  0.175 &  0.172 & &  0.148 &  0.107  \\
\noalign{\smallskip}\hline \noalign{\smallskip}
\end{tabular}
\end{table}

For 6250/4.0/$-4.0$, the MARCS and ATLAS9-ODFNEW model atmospheres provide consistent within 0.036~dex NLTE abundance corrections. Slightly larger differences of up to 0.058~dex are obtained for the strong lines, Ca\ione\ 4226~\AA\ and Ca\ii\ 8498~\AA, in the cool giant atmosphere. We remind that the MARCS models with $\logg \le 2$ were computed as spherically-symmetric, and the difference in temperature stratification between the spherically-symmetric and plane-parallel (ATLAS9-ODFNEW) models can explain, in part, differences in $\Delta_{\rm NLTE}$ for strong spectral lines.

\section{Comparisons with other studies}\label{Sect:others}

The NLTE methods based on comprehensive model atoms and the most up-to-date atomic data have been developed in the literature for many chemical species observed in spectra of the Sun and F-G-K type stars because the NLTE results are in demand in chemical abundance analyses of, in particular, VMP stars. For a common chemical species, the model atoms in different NLTE studies can differ by a treatment of inelastic collisions with electrons and hydrogen atoms and by the sources of transition probabilities and photoionization cross-sections. Different NLTE studies use different NLTE codes, with a different treatment of background opacity, and different model atmospheres. We compared our NLTE calculations with the NLTE abundance corrections from the other studies.   

\subsection{Lines of Fe\ione}

As shown in Fig.~\ref{Fig:compare1}, our results for lines of Fe\ione\ agree well with the NLTE abundance corrections from the NLTE\_MPIA database, which were computed using the model atom of \citet{Bergemann_fe_nlte} and the same treatment of collisions with H\ione, as in our calculations, namely, the formulas of \citet{Steenbock1984} with a scaling factor of \kH\ = 0.5. The differences in $\Delta_{\rm NLTE}$ between this study (TS) and NLTE\_MPIA mostly do not exceed 0.02~dex, with the maximal (TS -- NLTE\_MPIA) = 0.06~dex for Fe\ione\ 5506~\AA\ in the 6350/4.09/$-2.18$ model and Fe\ione\ 5041~\AA\ in the 4630/1.28/$-2.99$ model. 

\citet[][hereafter, Amarsi22]{2022A&A...668A..68A} provide the NLTE abundance corrections computed with the 1D and 3D model atmospheres. The 3D-NLTE calculations were performed for a limited atmospheric parameter range ($\Teff$ = 5000--6500~K, $\logg$ = 4.0 and 4.5, [Fe/H] = 0 to $-3$) and a limited number of Fe\ione\ lines. We selected Fe\ione\ 5232~\AA\ for a comparison. Amarsi22 computed more positive NLTE corrections compared with ours (Fig.~\ref{Fig:compare1}), by 0.07 to 0.27~dex in the 1D case and by 0.14 to 0.39~dex in the 3D case. The difference between 1D-NLTE corrections is most probably due to a different treatment of the Fe\ione\ + H\ione\  collisions in this and Amarsi22's studies. For H\ione\ impact excitation and charge transfer, Amarsi22 apply the asymptotic model of \citet{2018A&A...612A..90B} complemented by the free electron model of \citet{1991JPhB...24L.127K} for the b-b transitions. We showed earlier \citep{2019A&A...631A..43M} that compared with the \citet{Steenbock1984} formulas with \kH\ = 0.5 using data of \citet{2018A&A...612A..90B} leads to stronger NLTE effects. For example, $\Delta_{\rm NLTE}$ = 0.08~dex and 0.35~dex, respectively, for Fe\ione\ 5232~\AA\ in the 6350/4.09/$-2.18$ model atmosphere. In the 3D model atmospheres, the NLTE effects for Fe\ione\ are stronger than in the 1D models, and notable departures from LTE appear for lines of Fe\ii, in contrast to the 1D case, such that, for two benchmark VMP stars, Amarsi22 (see their Table~5) obtain similar abundance differences between Fe\ione\ and Fe\ii\ in the 1D-NLTE and 3D-NLTE calculations. To remind the reader, our 1D-NLTE approach for Fe\ione -\ii\ makes the spectroscopic distances of the VMP stellar sample to be consistent with the Gaia eDR3 ones (Sect.~\ref{sect:distances}).

\begin{figure*}  
 \begin{minipage}{150mm}

\parbox{0.3\linewidth}{\includegraphics[scale=0.45]{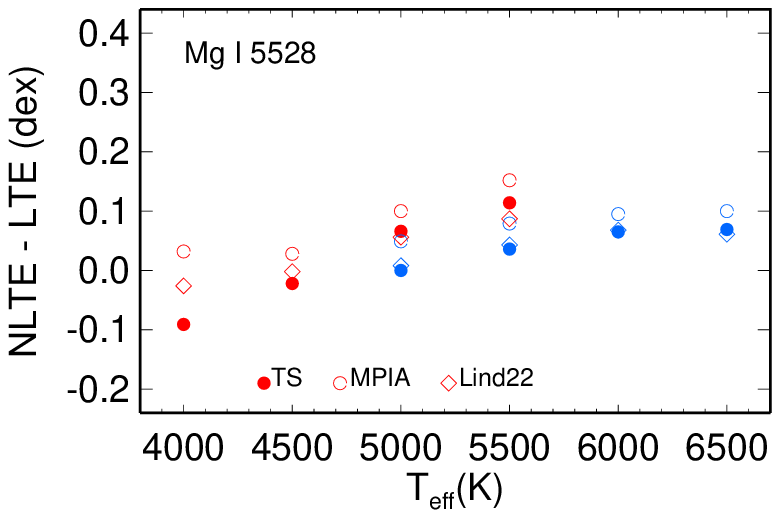}\\
\centering}
\hspace{15mm}
\parbox{0.3\linewidth}{\includegraphics[scale=0.45]{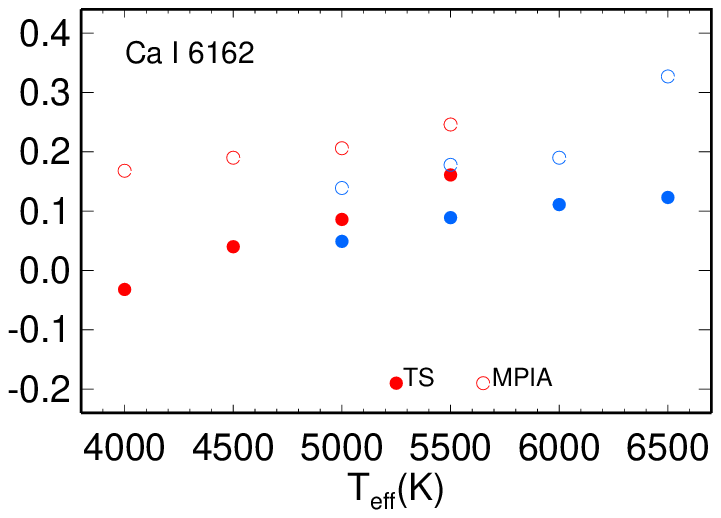}\\
\centering}

\vspace{-5mm}
\parbox{0.3\linewidth}{\includegraphics[scale=0.45]{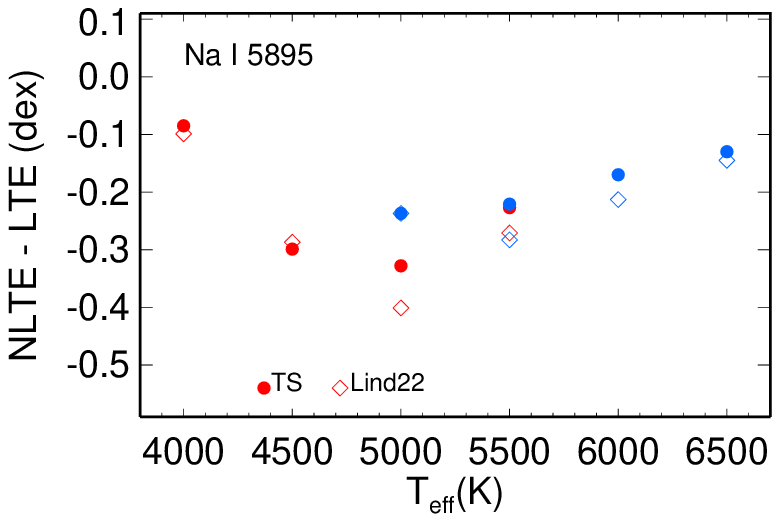}\\
\centering}
\hspace{15mm}
\parbox{0.3\linewidth}{\includegraphics[scale=0.45]{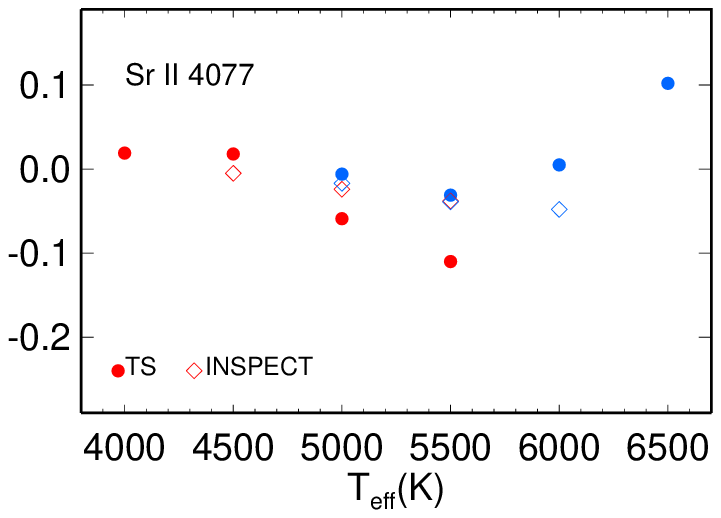}\\
\centering}
\hfill
\\[0ex]

\vspace{-5mm}
  \caption{\label{Fig:compare3} NLTE abundance corrections for Mg\ione\ 5528~\AA, Ca\ione\ 6162~\AA, Na\ione\ 5895~\AA, and Sr\ii\ 4077~\AA\ depending on $\Teff$ in the model atmospheres with [Fe/H] = $-3$ for two values of $\logg$: 2.5 (red symbols) and 4.0 (blue symbols). The data taken from \citet{2022A&A...665A..33L} and the INSPECT database are shown by open diamonds and from the NLTE\_MPIA database by open circles, while our calculations (TS) by filled circles.}
\end{minipage}
\end{figure*}

\subsection{Lines of Na\ione, Mg\ione, Ca\ione, Ca\ii, and Sr\ii}

We selected Mg\ione\ 5528~\AA, in order to compare our NLTE calculations with the 1D-NLTE corrections provided by the NLTE\_MPIA database and by \citet[][hereafter, Lind22]{2022A&A...665A..33L}. The used model atoms \citep[from][for NLTE\_MPIA]{2017ApJ...847...15B} are similar to ours, including a treatment of collisions with H\ione\ atoms. As seen in Fig.~\ref{Fig:compare3}, our calculations agree very well with those of Lind22. The differences in $\Delta_{\rm NLTE}$ do not exceed 0.01~dex and 0.02~dex for the $\logg$ = 4.0 and 2.5 models, respectively. The exception is the 4000/2.5/$-3$ model, for which we obtained a 0.065~dex more negative $\Delta_{\rm NLTE}$. NLTE\_MPIA provides more positive NLTE corrections compared with ours, by 0.03--0.05~dex. The difference is 0.12~dex for the 4000/2.5/$-3$ model.

Similar model atoms of Na\ione\ were used in this study and by Lind22. The differences in $\Delta_{\rm NLTE}$ for Na\ione\ 5895~\AA\ are very small ($\sim$0.01~dex) for the coolest and the hottest temperatures in Fig.~\ref{Fig:compare3}. It is difficult to explain why TS -- Lind22 = 0.07~dex for the 5000/2.5/$-3$ model, but TS -- Lind22 = 0.00 for 5000/4.0/$-3$.

For lines of Sr\ii, the 1D-NLTE corrections are provided by the INSPECT database. Their NLTE calculations were performed with the model atom developed by \citet{2012A&A...546A..90B} and did not take into account collisions with H\ione\ atoms. This is in contrast to this study based on quantum-mechanical rate coefficients for the Sr\ii\ + H\ione\ collisions. The atmospheric parameter range is narrower in INSPECT compared with this study, namely: 4400~K $\le \Teff \le$ 6400~K, 2.2 $\le \logg \le$ 4.6, $-3.9 \le$ [Fe/H] $\le 0$. The differences in $\Delta_{\rm NLTE}$ for Sr\ii\ 4077~\AA\ are small except the models 5500/2.5/$-3$ and 6000/4.0/$-3$, where TS -- INSPECT = $-0.07$~dex and +0.05~dex, respectively (Fig.~\ref{Fig:compare3}). 

The 1D-NLTE corrections for the Ca\ione\ lines at the NLTE\_MPIA database were computed with the model atom developed by \citet{mash_ca} and using \citet{Steenbock1984} formulas with \kH\ = 0.1 for calculating hydrogen collision rates. In this study, we applied the same model atom, however, the Ca\ione\ + H\ione\ collisions were treated using quantum-mechanical rate coefficients from \citet{2017ApJ...851...59B}. As seen in Fig.~\ref{Fig:compare3}, NLTE\_MPIA provides systematically greater NLTE corrections for Ca\ione\ 6162~\AA\ compared with our data, by 0.08 to 0.20~dex, probably due to a simplified treatment of hydrogenic collisions.

Ignoring the Ca\ii\ + H\ione\ collisions in the SE calculations resulted in stronger NLTE effects for the Ca\ii\ triplet lines in  \citet{2011MNRAS.418..863M} study compared with ours. For example, \citet{2011MNRAS.418..863M} report the NLTE/LTE equivalent ratios of 1.28 and 1.16 for Ca\ii\ 8498 and 8542~\AA, respectively, in the 4250/1.5/$-4.0$ model, while our corresponding values are 1.22 and 1.12.


\begin{figure}  
 \begin{minipage}{85mm}
\centering
	\includegraphics[width=0.75\textwidth, clip]{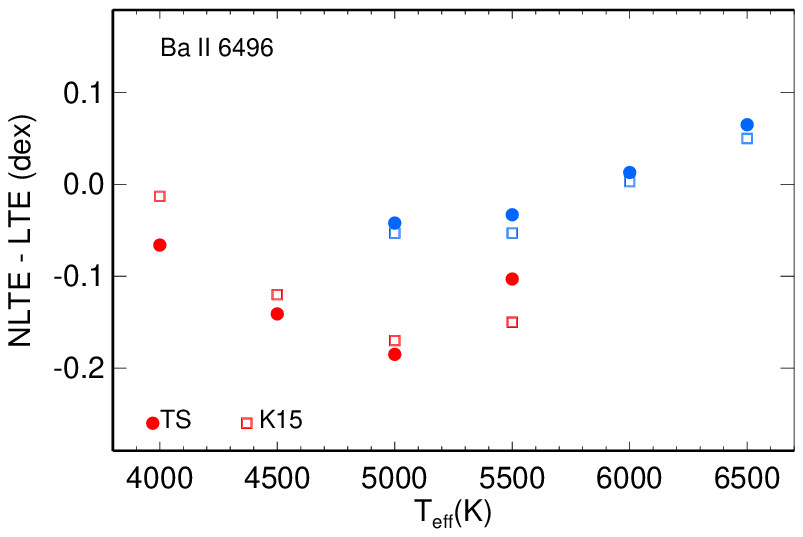}
  \caption{\label{Fig:compare_ba} NLTE abundance corrections for Ba\ii\ 6496\,\AA\ depending on $\Teff$ in the model atmospheres with [Fe/H] = $-2$ for two values of $\logg$: 2.5 (red symbols) and 4.0 (blue symbols). The filled circles and squares show our calculations (TS) and results of \citet{2015A&A...581A..70K}. }
\end{minipage}
\end{figure}

\subsection{Lines of Ba\ii}

Finally, we compared our results with the 1D-NLTE corrections calculated by \citet[][K15]{2015A&A...581A..70K} for lines of Ba\ii. \citet{2015A&A...581A..70K} provide the data for the $-2 \le$ [Fe/H] $\le 0.5$ metallicity range. Therefore, $\Delta_{\rm NLTE}$ comparisons are presented in Fig.~\ref{Fig:compare_ba} for the same temperatures and surface gravities, as in Fig.~\ref{Fig:compare3}, but for [Fe/H] = $-2$. The differences in $\Delta_{\rm NLTE}$ for Ba\ii\ 6496\,\AA\ do not exceed 0.02~dex except the coolest and the hottest giant atmospheres, where TS -- K15 = $-0.05$~dex and +0.05~dex, respectively.

To summarise this section, the situation with the 1D-NLTE corrections for lines of Na\ione, Mg\ione, and Fe\ione\ looks good. For each of these chemical species, there are, at least, two independent NLTE studies that predict consistent within 0.01-0.02~dex NLTE corrections and provide the grids which cover the full range of atmospheric parameters of VMP stars. For Sr\ii\ and Ba\ii, the NLTE corrections predicted by the independent studies agree reasonably well in the overlapping atmospheric parameter range.

\section{Final remarks}\label{Sect:conclusion}

This study presents grids of the 1D-NLTE abundance corrections for the Na\ione, Mg\ione, Ca\ione, Ca\ii, Ti\ii, Fe\ione, Zn\ione, Zn\ii, Sr\ii, and Ba\ii\ lines, which are used in the galactic archaeology research. The range of atmospheric parameters represents VMP stars on various evolutionary stages and covers 4000~K $\le \Teff\ \le$ 6500~K, 0.5 $\le \logg\ \le$ 5.0, and $-5.0 \le$ [Fe/H] $\le -2.0$. The NLTE corrections for Zn\ione, Zn\ii, Sr\ii, and Ba\ii\ have been calculated for the first time for such a broad atmospheric parameter range. Compared to the data available in the literature, our NLTE corrections for lines of Ca\ione, Ca\ii, Zn\ione, Zn\ii, Sr\ii, and Ba\ii\ are based on accurate treatment of collisions with H\ione\ atoms in the statistical equilibrium calculations.

In the same model atmosphere, the NLTE abundance corrections may have different magnitude and sign for lines of the same chemical species, for example $\Delta_{\rm NLTE}$ = 0.092~dex (Mg\ione\ 5528~\AA) and $\Delta_{\rm NLTE} = -0.083$~dex (Mg\ione\ 5172~\AA) in the 4500/1.5/$-3.5$ model. Accounting for the NLTE effects in stellar abundance determinations is expected to improve an accuracy of the obtained results.

In the same model atmosphere, the NLTE abundance corrections may have different magnitude and sign for lines of different chemical species, for example, $\Delta_{\rm NLTE}$ = $-0.222$~dex (Na\ione\ 5895~\AA) and $\Delta_{\rm NLTE}$ = 0.092~dex (Mg\ione\ 5528~\AA) in the 4500/1.5/$-3.5$ model. Therefore, an appropriate treatment of the line formation is obligatory for the studies based on analysis of the stellar element abundance patterns.

For all spectral lines and chemical species, the NLTE corrections depend on metallicity. Neglecting the NLTE effects in stellar abundance determinations leads to distorted galactic abundance trends and incorrect conclusions on the Galactic chemical evolution.

We show that, for common spectral lines and the same atmospheric parameters, independent NLTE studies of Na\ione, Mg\ione, and Fe\ione\ predict consistent 1D-NLTE abundance corrections, with the difference of 0.01-0.02~dex in $\Delta_{\rm NLTE}$.

The obtained results are publicly available. At the website INASAN\_NLTE ({\url{http://spectrum.inasan.ru/nLTE2/}}), we provide the tools for calculating online the NLTE abundance correction(s) for given line(s) and given atmospheric parameters.


\section*{Acknowledgements}
This research has made use of the data from the European Space Agency (ESA) mission Gaia\footnote{\url{https://www.cosmos.esa.int/gaia}}, processed by the Gaia Data Processing and Analysis Consortium (DPAC\footnote{\url{https://www.cosmos.esa.int/web/gaia/dpac/consortium}}). 
This research has made use of the MARCS and ADS\footnote{\url{http://adsabs.harvard.edu/abstract\_service.html}} databases. L.M. thanks the Russian Science Foundation (grant 23-12-00134) for a partial support of this study (Sections 1, 2, 4, 5). T.S. acknowledges a partial support (Section 3) from the MK project, grant 5127.2022.1.2.
 
\section{Data availability}

All our results are publicly available at the website INASAN\_NLTE ({\url{http://spectrum.inasan.ru/nLTE2/}}). 

\bibliography{atomic_data,mashonkina,nlte,references,nlte_corr}
\bibliographystyle{mnras}

\label{lastpage}
\end{document}